# Revisiting PbTe to identify how thermal conductivity is really limited

Shenghong Ju[1,2], Takuma Shiga[1], Lei Feng[1], Junichiro Shiomi[1,2,3,*]

[1]Department of Mechanical Engineering, The University of Tokyo, 7-3-1 Hongo, Bunkyo, Tokyo 113-8656, Japan

[2]Center for Materials research by Information Integration, National Institute for Materials Science (NIMS), 1-2-1 Sengen, Tsukuba, Ibaraki 305-0047, Japan

[3]Core Research for Evolutional Science and Technology (CREST), Japan Science and Technology Agency (JST), 4-1-8, Kawaguchi, Saitama 332-0012, Japan

*Corresponding email: shiomi@photon.t.u-tokyo.ac.jp

## ABSTRACT

Due to the long range interaction in lead telluride (PbTe), the transverse optical (TO) phonon becomes soft around the Brillouin zone center. Previous studies have postulated that this zone-center softening causes the low thermal conductivity of PbTe through either enlarged phonon scattering phase space and/or strengthened lattice anharmonicity. In this work, we reported an extensive sensitivity analysis of the PbTe thermal conductivity to various factors: range and magnitude of harmonic and anharmonic interatomic force constants, and phonon wavevectors in the three-phonon scattering processes. The analysis reveals that the softening by long range harmonic interaction itself does not reduce thermal conductivity and it is the large magnitude of the anharmonic (cubic) force constants that realizes low thermal conductivity, however, not through the TO phonons around the zone center but dominantly through the ones with larger wavevectors in the middle of Brillion zone. The work clarifies that local band softening cannot be a direct finger print for low thermal conductivity and the entire Brillion zone needs to be characterized on exploring low thermal conductivity materials.

## I. INTRODUCTION

Thermal transport in crystals with strong lattice anharmonicity has attracted great attention as a source of anomalous high-order lattice dynamics and means to realize thermoelectrics with low intrinsic thermal conductivity. Lead telluride (PbTe) is the most studied material, owning to its high thermoelectric performance [1] and high structural-symmetry enabling in-depth analysis and detailed experimental characterization [2]. The measured bulk thermal conductivity of single crystal PbTe is around 2.2 $Wm^{-1}K^{-1}$ [3-5] at room temperature. This is extremely low compared with other high symmetry thermoelectric crystals that, to achieve similar thermal conductivity, need to be nanostructured in forms of alloys [6,7], superlattices [8,9] and nanocrystals [10,11].

Over the last decade, there has been a great advance in associating this low thermal conductivity to anharmonic lattice dynamics characteristics. The transverse optical (TO) phonon at Γ point is extremely soft and has relatively low frequency [12-16], and exhibit diverging Grüneisen-parameter [14,17]. The high sensitivity of the Γ-TO mode to the volume change suggested large coupling between the TO modes and longitudinal acoustic (LA) modes [12], and the intrinsic scattering of LA modes would pull down the thermal conductivity. This was later quantified by first-principles anharmonic lattice dynamics [17], which showed that TO phonons in overall is responsible for 61.5% of the scattering rates of LA phonons whose contribution to thermal conductivity is consequently limited to 21%. Detailed inelastic neutron scattering also supports the anharmonic TO-LA interaction through broadened line shapes and avoided crossing [2].

An important point here is that anharmonic features including non-Gaussian radial distribution function (RDF) in neutron diffraction [18] and double peaks in neutron scattering [2,19,20] can be reproduced by first-principles anharmonic lattice dynamics method, and thus they can be captured by perturbation theory [21]. The perturbation analysis has revealed an important role of certain interatomic force constants (IFCs). Non-Gaussian RDF and the double-peak line shape mainly originate from the nearest-neighbor cubic IFCs in the [100] direction, whose magnitude is considerably larger than other cubic IFCs. The softening of TO modes was attributed to long range harmonic IFCs in the [100] direction with large magnitude due to the resonant bonding: introduced by long-range polarization of *p*-electron distribution to atomic displacement, which was found to be a common feature in lead chalcogenides [22].

Besides the above, it has been shown that, as temperature increases and the TO mode stiffens, the phase space for scattering between the TO and acoustic phonons disappears, which weakens the temperature dependence of thermal conductivity [23]. Anomalous splitting of the spectrum at the zone center captured by both inelastic neutron scattering measurements [2] and first-principles calculations [20] was found to be related with the nearest neighbor cubic and quartic anharmonic interaction. There has also been an experimental reports showing that the lead ion in PbTe and PbS is on average displaced by 0.2 Å from the rock-salt lattice position, which can cause phonon scattering [24].

As above, a good progress has been made to characterize and understand the origin of specific features of harmonic and anharmonic lattice dynamics, however, how that really contributes to low thermal conductivity remains vague. Soft zone-center TO mode with large Grüneisen number has been suggested to play the core role, but it is not clear at this point, whether the role is to increase the scattering phase space or to increase the scattering magnitude. Moreover, there is a question whether the softening of the modes around zone-center would really have a significant impact to board range of modes in the Brillion zone that contribute to thermal conductivity. Clarifying these points is important as it would influence the strategy in looking for low thermal conductivity materials. For this, in this work, we conduct a sensitivity study of PbTe thermal conductivity with respect to participation of specific range of IFCs and phonon modes in the entire first Brillion zone to identify their contribution to thermal conductivity.

## II. METHODOLOGY

The calculations are based on lattice dynamics using IFCs obtained from first-principles. We adopted 4×4×4 conventional supercell containing 512 atoms, which is larger than any systems used in previous calculations [17,21,22,25], to make sure the calculated system fully captures the long range interaction. Both harmonic and anharmonic IFCs were calculated by density functional theory (DFT) and the real-space displacement method [26,27], using Quantum ESPRESSO [28] and ALAMODE [29] packages. Including spin-orbit coupling (SOC) leads to softening of all phonon modes and improves the agreement with experimental dispersion relations [30]. The effect of SOC on thermal conductivity calculation has been discussed in the previous work [25]. By considering the SOC effect, the relaxation time of all phonon modes are noticeably larger, which results in larger thermal conductivity than that without SOC effect at room temperature. In this work, we used the Perdew-Zunger norm conserving pseudopotentials incorporating relativistic effects [31], which includes the SOC effect under the local density approximation (LDA) for electron exchange-correlation potential. The energy cutoff of 60 Ryd were used for DFT calculations, and the lattice constant after total energy minimization is 6.289 Å. The *k*-mesh used in the calculation is 2×2×2 for the conventional supercell. The effect of ionic charges in PbTe was considered by performing density functional perturbation theory (DFPT) calculation [30-32]. The obtained phonon dispersion of PbTe along high symmetry lines shown in Fig. 1(a). Although TO frequency at zone center is larger than that of experiment [15] by about 0.4 THz, which could be due to the use of ground-state harmonic force constants, the overall feature agrees reasonably well with the experimental results [15] as well as previous first-principle calculations [22]. We also performed the calculations for different sizes of supercell using the supercell approach (see Fig. S1 and S2 in Supplementary Material [33]), it is shown that the dispersion relations and the TO frequency at zone center approach the DFPT results with increasing size of supercell. One drawback of the supercell approach is that large supercells are needed to calculate the IFCs accurately. In this work, we adopted 4×4×4 supercell, which is the maximum size accessible computers could afford, to capture the long range lattice dynamics in PbTe. Particularly, we compared the TO phonon dispersion along [100] direction with that of Lee *et al* [22] by turning off the 4th and 8th neighbor harmonic force constants, which results in a similar non-dispersive TO branch. The minor difference in amplitude possibly comes from the difference in the pseudopotentials used.

The Boltzmann transport equation (BTE) with relaxation time approximation (RTA) [26,34,35] was employed to calculate the lattice thermal conductivity,

$$\kappa_{\alpha\beta} = \frac{1}{\Omega N_q} \sum_{\mathbf{q},j} c_{\mathbf{q}j} v_{\mathbf{q}j}^{\alpha} v_{\mathbf{q}j}^{\beta} \tau_{\mathbf{q}j}, \tag{1}$$

where $\Omega$ is the volume of the primitive unit cell, $N_q$ is the number of $q$ points, $\alpha$ and $\beta$ indicate the velocity components, $c_{\mathbf{q}j}$, $v_{\mathbf{q}j}$ and $\tau_{\mathbf{q}j}$ are heat capacity, group velocity and relaxation time of the phonon with wave vector $\boldsymbol{q}$ and mode $j$. The phonon relaxation time $\tau_q$ is further given by,

$$\tau_q^{-1} = \frac{\pi}{N_q \hbar^2} \sum_{\boldsymbol{q}',\boldsymbol{q}''} \left| V^{(3)}(-\boldsymbol{q};\boldsymbol{q}';\boldsymbol{q}'') \right|^2 \Big[ (\bar{n}_{\boldsymbol{q}'} + \bar{n}_{\boldsymbol{q}''} + 1)\delta(\omega_{\boldsymbol{q}} - \omega_{\boldsymbol{q}'} - \omega_{\boldsymbol{q}''}) + 2(\bar{n}_{\boldsymbol{q}'} - \bar{n}_{\boldsymbol{q}''})\delta(\omega_{\boldsymbol{q}} + \omega_{\boldsymbol{q}'} - \omega_{\boldsymbol{q}''}) \Big], \tag{2}$$

where $V^{(3)}$ indicates the anharmonicity magnitude of three-phonon scattering, $\bar{n}$ is the Bose-Einstein distribution, $\boldsymbol{q}$, $\boldsymbol{q}$' and $\boldsymbol{q}$" are triplet pairs of three-phonon scattering which satisfy the energy and momentum conservation. The tetrahedron method [36] was used for Brillouin zone integration with 20×20×20 uniform $q$ mesh. Figure 1(b) shows the calculated thermal conductivity at room temperature using different harmonic and cubic force cutoff distances. For 4×4×4 conventional supercell system, the longest interaction can reach the 31th neighbor. Note that as heat conduction in PbTe is isotropic, thermal conductivity is quantified as a scalar value.

With increasing cubic cutoff, the thermal conductivity decreases and converges to almost a constant value of 3.1 $Wm^{-1}K^{-1}$. When fixing the cubic cutoff at the 6th neighbor and increasing the harmonic force cutoff gradually, the thermal conductivity keeps increasing and finally converges to a constant value at 28th neighbor. For the commonly used 4×4×4 primitive supercell system the reported thermal conductivity is 2.1 $Wm^{-1}K^{-1}$ [17,22,25] and the corresponding longest interaction length is shorter than the 28th neighbor. This indicates that when calculating the phonon transport in crystals with long range force interaction, the system size and cutoff length should be carefully chosen, otherwise the calculated thermal conductivity will be under- or over-estimated. Consequently, the saturated value in the current calculation becomes larger than that in the experiment, which could be due to small but non-negligible disorders in the experiment or the selection of the exchange-correlation functions and pseudo-potentials in the calculation. The quasi-harmonic approximation on calculating the dispersion relations instead of self-consistently including anharmonicity may also be a potential source of uncertainty [37]. The point made here is that, although it is possible to post-tune/reason the calculated values to match the experiments, that should not be done at the expense of missing key physical component, and for the purpose of this work, the somewhat larger value is acceptable.

## III. RESULTS AND DISCUSSION

To study sensitivity of thermal conductivity on the range of resonant bonds, we manipulate the long range harmonic (4th and 8th neighbor interaction) and anharmonic cubic (the largest first-nearest-neighbor IFCs along [100] direction: $\psi_{\mathrm{PbPbTe}}^{\mathrm{xxx}}$ and $\psi_{\mathrm{PbTeTe}}^{\mathrm{xxx}}$, which are only 2 components out of 633 irreducible cubic IFCs) interaction by scaling the IFCs, as shown in Fig. 2(a). The obtained responses of thermal conductivity to harmonic and cubic IFCs are quite opposite as shown in Fig. 2(b). The thermal conductivity increases by increasing the range harmonic interaction, and in contrast, it decreases with increasing the range cubic interaction (see Fig. S3 in Supplementary Material [33] for similar result when scaling all the components of 1st and 4th neighbor cubic IFCs). The calculated phonon dispersion relations for typical scaling factors of the 4th and 8th neighbor harmonic interaction are shown in Fig. 3(a). With increasing the scaling factor, the strength of long range interaction increases, and the corresponding TO branches become more dispersive. This trend agrees with the results reported by Lee *et al* [22], who suggested that softened TO mode around Γ point reduces thermal conductivity due to enlarging both scattering phase space and anharmonic scattering. However, our result shows that with softer TO phonons at Γ point, the thermal conductivity of PbTe in fact increases.

To explore and clarify the mechanism behind, the three-phonon scattering phase space (SPS) [38], which describes the amount of phonon scattering channels, is calculated as

$$P_3(\boldsymbol{q}) = \frac{1}{3m^3}\left(2P_3^{(+)}(\boldsymbol{q}) + P_3^{(-)}(\boldsymbol{q})\right), \tag{3}$$

where $m$ is the number of phonon branches, and

$$P_3^{(\pm)}(\boldsymbol{q}) = \frac{1}{N_q}\sum_{\boldsymbol{q}',\boldsymbol{q}''}\delta\left(\omega_{\boldsymbol{q}} \pm \omega_{\boldsymbol{q}'} - \omega_{\boldsymbol{q}''}\right)\delta_{\boldsymbol{q}\pm\boldsymbol{q}',\boldsymbol{q}''+G}\,. \tag{4}$$

The total SPS is then given by,

$$P_3 = \frac{1}{N_q}\sum_{q} P_3(\boldsymbol{q})\,. \tag{5}$$

The calculated total SPS and $\Gamma_{TO}$-SPS are shown in Fig. 3(b). $\Gamma_{TO}$-SPS here denotes the partial SPS (subset of $P_3$) of scattering processes involving the TO phonon at the zone center (Γ). The total SPS first decreases to a minimum value at scaling factor of 0.3 and then increases to an almost constant value for scaling factors over 0.5. The $\Gamma_{TO}$-SPS exhibits similar but opposite trend. Within the range of scaling factors, the thermal conductivity increases monotonically but the SPS shows more complicated trend, and there is no correlation between SPS and the thermal conductivity.

The plots of mode-dependent thermal conductivity, phonon mean free path (MFP),

relaxation time, and group velocity as a function of phonon frequency (see Fig. S4 in Supplementary Material [33]) are widely used for discussing thermal conductivity from phonon kinetics viewpoint, but since comparing such plots for different scaling factors makes the differences visually unclear due to the fluctuation (i.e. mode-dependence), here we instead plot the average group velocity versus the scaling factor in Fig. 3(b). With increasing long range harmonic interaction, the phonon group velocity clearly increases due to more dispersive TO and acoustic phonon branches, and therefore, thermal conductivity increases. This indicates that the long range harmonic interaction is not the cause of the low thermal conductivity.

To further explore the mechanism behind the low thermal conductivity of PbTe, here we propose an analysis method to quantitatively evaluate the contribution of different three-phonon scattering processes to heat conduction by removing specific scattering processes as shown in Fig. 4. The thermal conductivity enhancement after removing scattering of different combinations of three phonons is shown in Fig. 5 (a). The result indicates that A-A-O processes (i.e. either creation or annihilation among two acoustic (A) phonons and one optical (O) phonon) dominates the heat conduction in PbTe, the next two important processes are A-O-O and A-A-A. In fact, this order of significance is the same as that of partial SPS of specific three-phonon combination reported previously [22]. However, the current analysis is different from the partial SPS calculation since it also incorporates magnitude of anharmonicity, and we clarify in the following that SPS characteristics is not the reason why these processes impact the thermal conductivity. Figure 5(a) also shows more detailed picture by breaking the processes into those of different polarization, where LA-TA-TO, LA-TA-TA, TA-TA-TO, TA-LO-TO, TA-TO-TO, LA-TA-LO are found to be the top six processes impacting the thermal conductivity. Four out of six processes involve TO phonons, and thus TO phonons do play a leading role, however, as we will also show in the following, those optical phonons are hardly near-zone-center modes.

The key here is to look into wavevector of phonons that impact scattering properties and thermal conductivity. This was done by removing three-phonon scattering processes that involve optical phonons with wave vectors of a specific absolute value $k_0=|\boldsymbol{k}|$ Fig. 4 (b), and evaluating the resulting variation in the properties. This sensitivity analysis then reveals $k_0$ of phonons that dominantly influence the properties. Figure 5 (b)-(c) show the variation in thermal conductivity ($\kappa$), SPS, relaxation time ($\tau$), and anharmonic amplitude (estimated by $V_3 \approx 1/(\tau P_3)$) when removing A-A-O and A-O-O scattering processes involving $k_0$ phonon. In practice, the process is removed if it involves at least one optical phonon with $|\boldsymbol{k}|=k_0$. The result of $\kappa$ in Fig. 5(b) shows that TO phonons near Γ point play small role and it is in fact the phonons with intermediate wavevectors that contribute the most to scattering. For A-A-O process, the value of dominant $k_0$ normalized by the zone boundary is 0.6, and for A-O-O process it is 0.3. The difference in the peak $k_0$ values between A-A-O and A-O-O processes is understandable, as energy conservation limits A-A-O processes to A+A↔O and A-O-O processes to A+O↔O, and thus, considering the momentum conservation, the optical phonons in the former process ends up with larger wavevectors whereas one of the two optical phonons in the latter can have relatively smaller wavevector. Nevertheless, none of the peak $k_0$ values are near Γ point, which suggests that the phonon scattering in the entire Brillion zone needs to be characterized on exploring for low thermal conductivity materials.

The profile of $\tau$ in Fig. 5(c) resembles that of $\kappa$, which is reasonable because the harmonic properties and thus group velocities are unchanged during this analysis. Now the issue is whether the variation in relaxation time comes from SPS or anharmonic amplitude ($V_3$), and it is clear by comparing the profiles in terms of the peak position and the amplitude that anharmonic amplitude exhibits much stronger correlation. The stronger correlation is much clearer when further breaking the scattering process by distinguishing phonon polarization as shown in Fig. 5 (d) and (e).

**IV. CONCLUSIONS**

In summary, we have performed a systematic sensitivity study of PbTe thermal conductivity with respect to the range/magnitude of harmonic and anharmonic interatomic force constants. By increasing the range of harmonic interaction, although TO phonons become softer at Γ point, the thermal conductivity of PbTe increases in contract to what has been previously postulated, due to increase in dispersion and thus group velocity. Sensitivity analysis to the largest nearest neighbor cubic force constants shows the low thermal conductivity is due to their large anharmonicity. Further sensitivity analysis incorporating phonon polarization and wavevector identifies that it originates from the magnitude of the anharmonic force constants, not through the TO phonons around the zone center, but dominantly through the larger wavevector TO phonons in the middle of the Brillion zone. The new understanding clarifies that the zone centered TO features such as band softening and Grüneisen divergence cannot be used as a direct finger print for low thermal conductivity and the entire Brillion zone needs to be characterized on exploring low thermal conductivity materials.

## ACKNOWLEDGMENTS

The authors thank the useful discussions with Keivan Esfarjani and Terumasa Tadano. The calculations in this work were performed using supercomputer facilities of the Institute for Solid State Physics, the University of Tokyo. This work was supported in part by "Materials research by Information Integration" Initiative ($MI^2I$) project and CREST Grant No. JPMJCR16Q5 from Japan Science and Technology Agency (JST), and KAKENHI Grants No. 16H04274 from Japan Society for the Promotion of Science (JSPS).

## Figures and captions

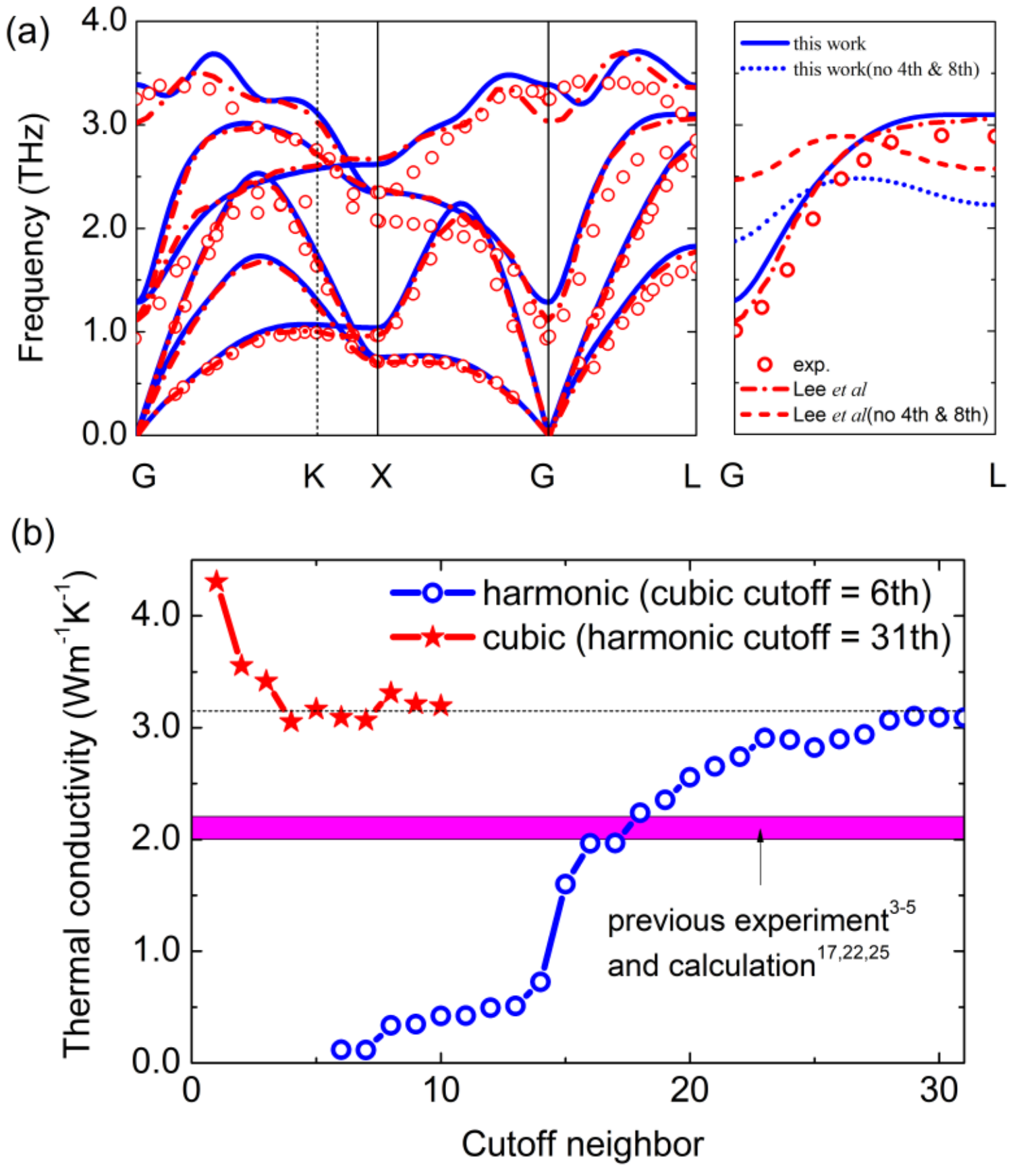

**Fig. 1** (a) Phonon dispersion of PbTe from first-principles calculation in this work (solid line), Lee *et al* [22] (dash-dot line), and experiments [15] (open circles). TO phonon dispersion along [100] direction with and without the long range of 4th and 8th neighbor harmonic force constants is also shown, where solid and dot lines are from current calculations, dash-dot and dash lines are from Ref. [22]. (b) Thermal conductivity of PbTe at room temperature versus the cutoff distance of harmonic and cuibc forces.

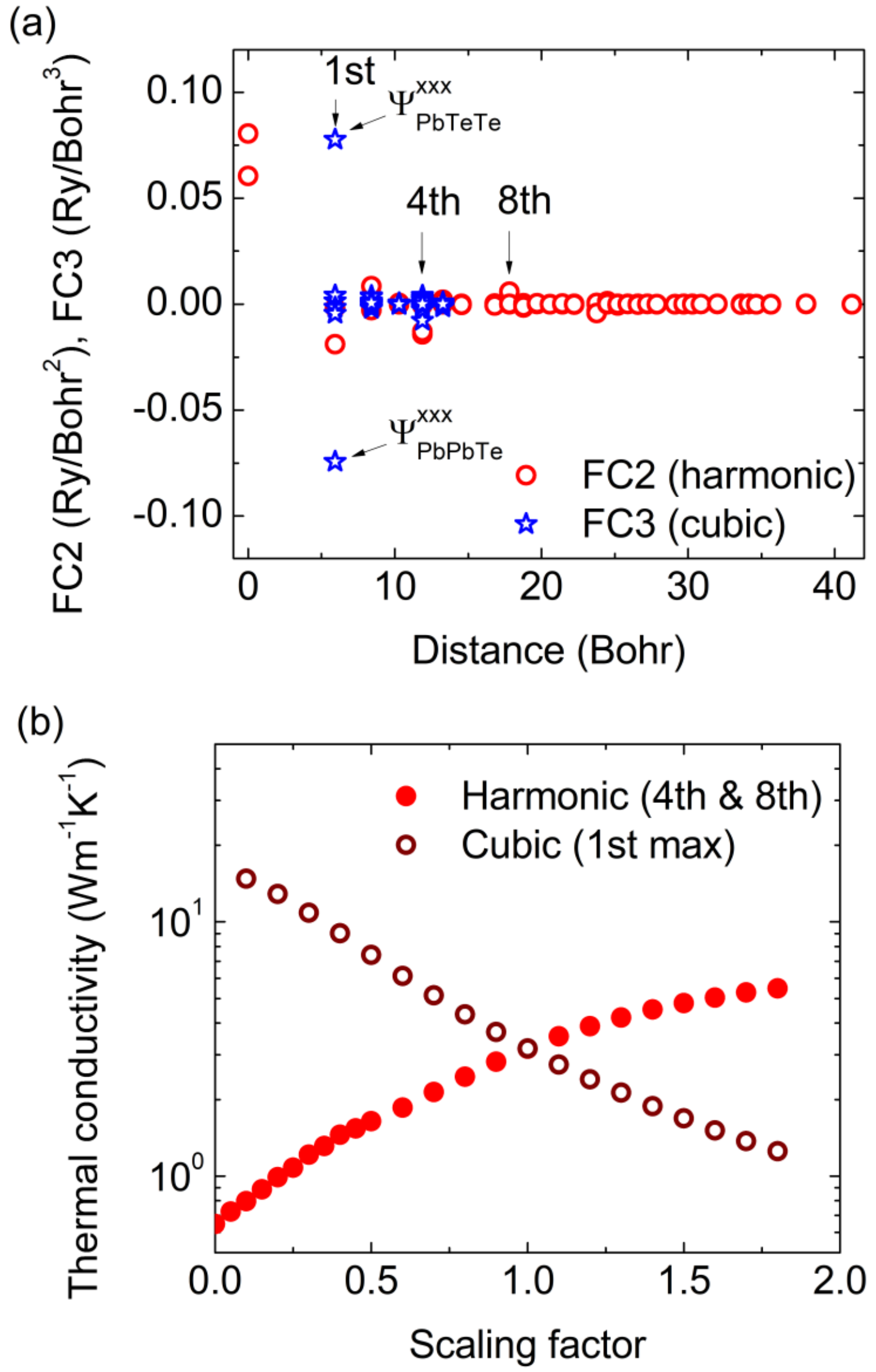

**Fig. 2** (a) Harmonic force constants (FC2) and cubic force constants (FC3) versus the neighbor distance. (b) Thermal conductivity versus the scaling factors for FC2 (4th and 8th neighbor interaction) and FC3 (the largest first-nearest-neighbor interaction along [100] direction: $\psi^{xxx}_{PbPbTe}$ and $\psi^{xxx}_{PbTeTe}$).

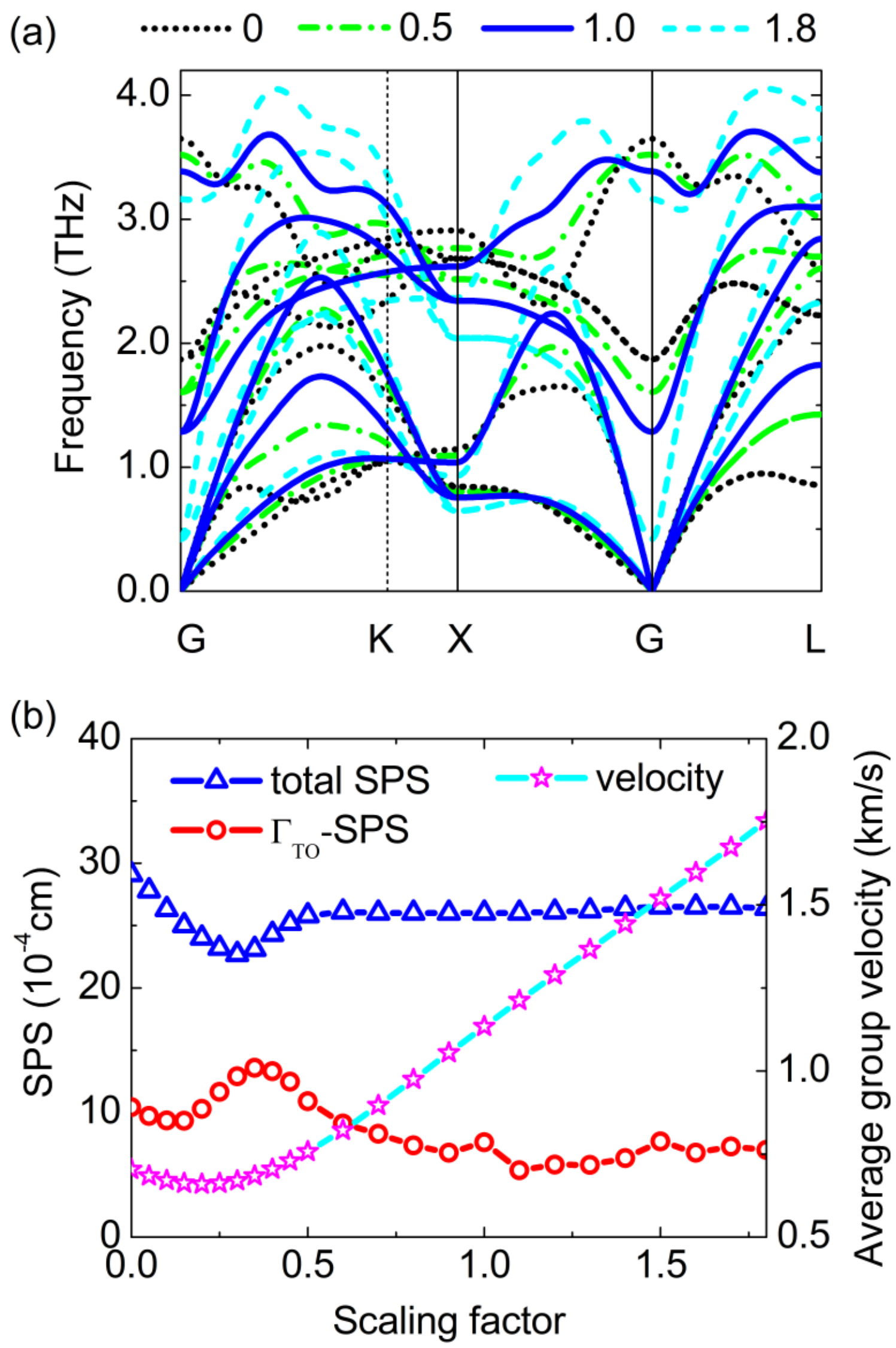

**Fig. 3** (a) Phonon dispersion of PbTe with different scaling factors of the 4th and 8th neighbor harmonic force constants. (b) The total and $\Gamma_{TO}$ phonon scattering phase space and average group velocity versus the scaling factors.

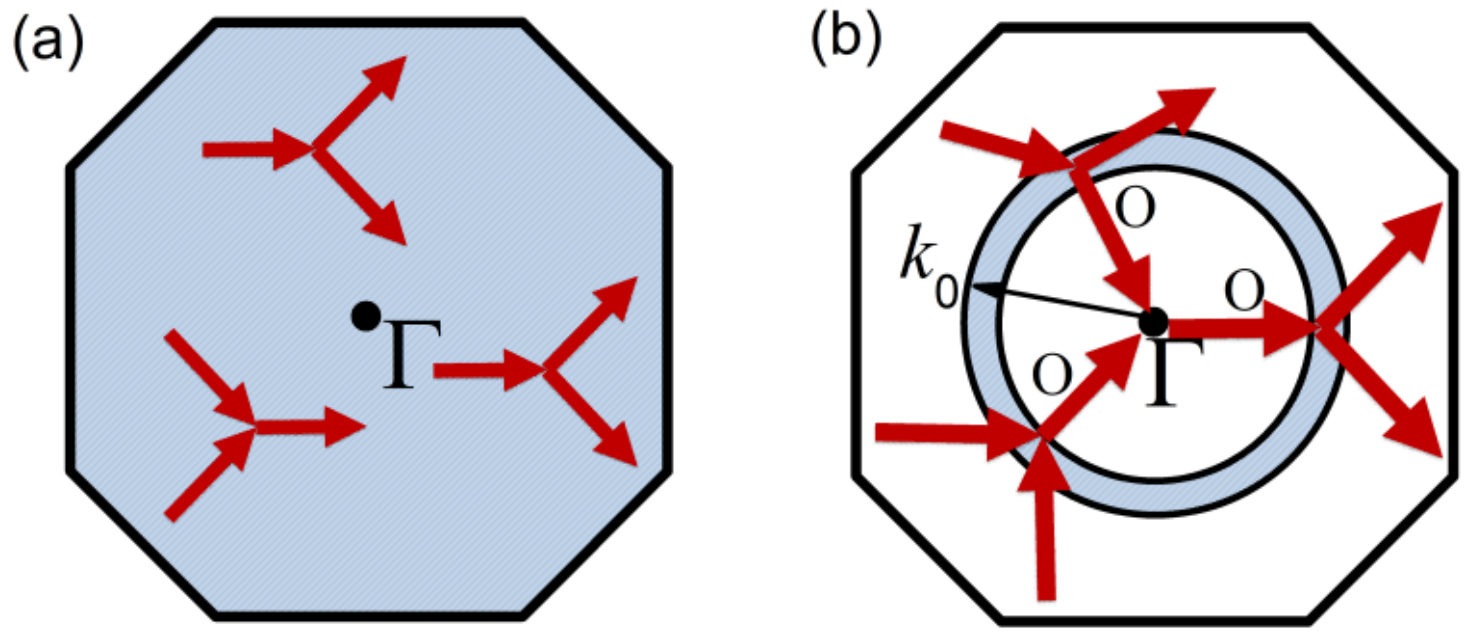

**Fig. 4** Sketch of removing specific three-phonon scattering processes: (a) removing scattering of different combinations of three phonons from entire zone, (b) removing three-phonon scattering processes that involve optical phonons with wave vectors of a specific absolute value $k_0=|\mathbf{k}|$.

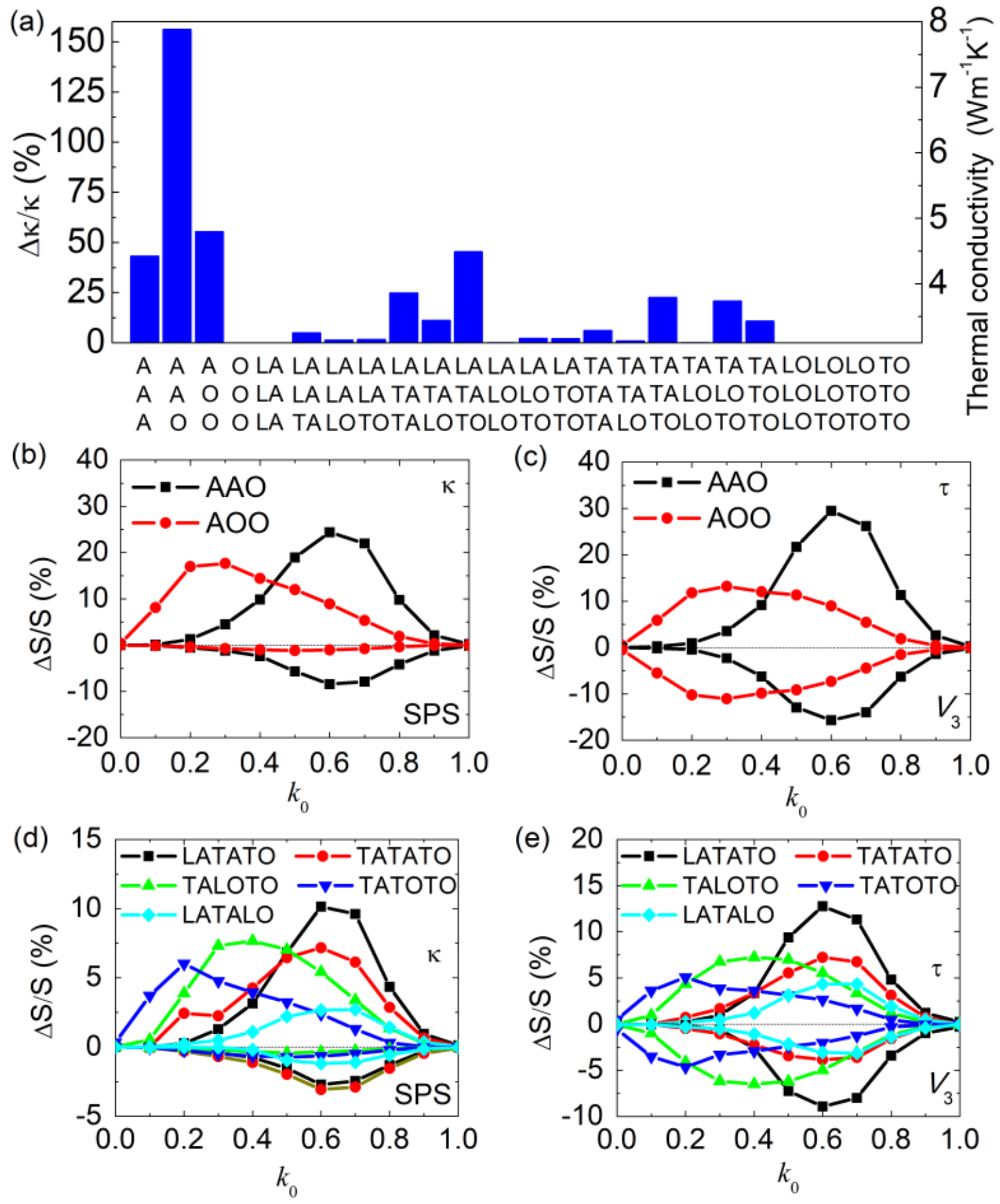

**Fig. 5** (a) Thermal conductivity enhancement of PbTe after removing specific three-phonon scattering process (b)-(e) The variation of thermal conductivity ($\kappa$), scattering phase space (SPS), relaxation time ($\tau$), and anharmonic amplitude ($V_3$) when removing scattering process involving optical phonons with absolute wavevector of $k_0$.